\documentstyle[sprocl]{article}

\def\be{\begin{equation}}
\def\ee{\end{equation}}
\def\bea{\begin{eqnarray}}
\def\eea{\end{eqnarray}}

\newcommand{\sss}{\smallskip}
\newcommand{\nin}{\noindent}

\begin{document}

\title{Dynamics and topology of the gauge-invariant gauge field in two-color QCD}

\author{Kurt Haller}

\address{Department of Physics, University of Connecticut, Storrs, CT 06269}

\maketitle\abstracts{A nonlinear integral equation that is responsible for the implementation of 
the non-Abelian Gauss's law is applied to an investigation of the topological features 
of two-color QCD and to a discussion of
their relation to QCD dynamics. We also draw a parallel between the nonuniqueness of the solutions 
of the equations that govern the gauge-invariant gauge field and Gribov copies.}

\section{Introduction}\label{sec:intro}
The investigation I am reporting concerns the implications of the constraints that apply in QCD
--- in particular the non-Abelian Gauss's law --- not only for the dynamics but also for the 
topological features of the theory.
Previous work has dealt with the implementation of Gauss's law, which we have carried out 
by constructing
states that are annihilated by the ``Gauss's law operator'' ${\hat {\cal G}}^{a}({\bf r})$
for the temporal ($A^a_0=0$) gauge,\cite{CBH2} where
\begin{equation}
{\hat {\cal G}}^{a}({\bf r})=\partial_{j}
\Pi_{j}^{a}({\bf r})+g{\epsilon}^{abc}A_{j}^{b}({\bf r})
\Pi_{j}^{c}({\bf r})+j_{0}^{a}({\bf r})\;\;
\mbox{and}\;\;j^a_{0}({\bf{r}})=
g\,\,\psi^\dagger({\bf{r}})\,
{\textstyle\frac{\tau^a}{2}}\,\psi({\bf{r}})
\label{eq:quark}
\end{equation}
and where $\Pi_{j}^{a}({\bf r})$ is the momentum conjugate to the gauge field 
$A_{j}^{a}({\bf r})$ as well as the negative chromoelectric field.
We have, furthermore, constructed gauge-invariant quark and gluon  operator-valued  fields 
and have transformed the QCD Hamiltonian into a form in which it is expressed 
entirely in terms of these gauge-invariant fields.\cite{BCH3,CHQC} 
An operator-valued field --- the so-called ``resolvent field'' 
$\overline{{\cal{A}}_{i}^{\gamma}}({\bf{r}})$,  
that is a non-local functional of the gauge field in the temporal gauge --- has a central role in these
developments. 
In the two-color SU(2) version of QCD, with which we will be concerned in this work, 
the resolvent field appears in the gauge-invariant gluon field $A_{{\sf GI}\,i}^{b}({\bf{r}})$ as shown by
\be
[\,A_{{\sf GI}\,i}^{b}({\bf{r}})\,{\textstyle\frac{\tau^b}{2}}\,]
=V_{\cal{C}}({\bf{r}})\,[\,A_{i}^b({\bf{r}})\,
{\textstyle\frac{\tau^b}{2}}\,]\,
V_{\cal{C}}^{-1}({\bf{r}})
+{\textstyle\frac{i}{g}}\,V_{\cal{C}}({\bf{r}})\,
\partial_{i}V_{\cal{C}}^{-1}({\bf{r}})\;.
\label{eq:AdressedAxz}
\end{equation}
\be
\mbox{where}\;\;\;\;V_{\cal{C}}({\bf{r}})=
\exp\left(\,-ig{\overline{{\cal{Y}}^\alpha}}({\bf{r}})
{\textstyle\frac{\tau^\alpha}{2}}\,\right)\,
\exp\left(-ig{\cal X}^\alpha({\bf{r}})
{\textstyle\frac{\tau^\alpha}{2}}\right)
\label{eq:el1}
\end{equation}
\begin{equation}
\mbox{and}\;\;\;\;V_{\cal{C}}^{-1}({\bf{r}})=
\exp\left(ig{\cal X}^\alpha({\bf{r}})
{\textstyle\frac{\tau^\alpha}{2}}\right)\,
\exp\left(\,ig{\overline{{\cal{Y}}^\alpha}}({\bf{r}})
{\textstyle\frac{\tau^\alpha}{2}}\,\right)
\label{eq:eldagq1}
\end{equation} 
\begin{equation}
\mbox{with}\;\;\;\;{\cal{X}}^\alpha({\bf{r}}) =
{\textstyle\frac{\partial_i}{\partial^2}}A_i^\alpha({\bf{r}})\;\;\;\;\mbox{and}\;\;\;\;
\overline{{\cal Y}^{\alpha}}({\bf r})=
{\textstyle \frac{\partial_{j}}{\partial^{2}}
\overline{{\cal A}_{j}^{\alpha}}({\bf r})}.
\label{eq:defXY}
\end{equation}
The gauge-invariant field $A_{{\sf GI}\,i}^b({\bf{r}})$ given in Eq. (\ref{eq:AdressedAxz})
can also be expressed as
\begin{equation}
A_{{\sf GI}\,i}^b({\bf{r}})=
A_{T\,i}^b({\bf{r}}) +
[\delta_{ij}-{\textstyle\frac{\partial_{i}\partial_j}
{\partial^2}}]\overline{{\cal{A}}^b_{j}}({\bf{r}}),
\label{eq:agidefin}
\ee
where $A_{T\,i}^b({\bf{r}})$ is the transverse part of the gauge field $A_i^b({\bf{r}})$ 
in the temporal gauge. The resolvent field is also required for defining  the
 gauge-invariant quark field 
\begin{equation}
{\psi}_{\sf GI}({\bf{r}})=V_{\cal{C}}({\bf{r}})\,\psi ({\bf{r}})
\;\;\;\mbox{\small and}\;\;\;
{\psi}_{\sf GI}^\dagger({\bf{r}})=
\psi^\dagger({\bf{r}})\,V_{\cal{C}}^{-1}({\bf{r}})\;.
\label{eq:psiqcdg1}
\end{equation}
We have shown that gauge transformations transform $V_{\cal{C}}({\bf{r}})$
in such a way, that they exactly compensate for the transformations undergone by 
$A_i^a({\bf{r}})$ and $\psi({\bf{r}})$ so that 
$A_{{\sf GI}\,i}^b({\bf{r}})$ and ${\psi}_{\sf GI}({\bf{r}})$ remain untransformed by any gauge
transformations.\cite{BCH3,CHQC}  \sss

Central to the implementation of Gauss's law and the construction of gauge-invariant operator-valued fields,
is a nonlinear integral equation that appeared in our earlier work as the defining 
equation for the resolvent field.\cite{CBH2}
In this nonlinear integral equation, functionals of the gauge field  appear as 
inhomogeneous source terms.  The significance of this integral equation is made manifest by
  a proof --- the so-called ``fundamental theorem'' --- that 
resolvent fields that solve this integral equation generate the gauge-invariant fields 
given in Eqs. (\ref{eq:AdressedAxz})-(\ref{eq:psiqcdg1}) as well as the states
that implement Gauss's law. 

\section{Topological configurations of the resolvent field}\label{sec:top}
At the QCD Workshop that met in Paris last year, I discussed the transformation of the QCD Hamiltonian into
a form that is expressed entirely in terms of gauge-invariant fields.\cite{CHQC,QCD4} 
In this form, the QCD Hamiltonian is particularly suited for applications to low-energy
phenomena, and displays quark-quark and quark-gluon interactions that are non-Abelian analogs of the 
Coulomb interaction in QED.
More recently, we have been 
studying the topological implications of the equations that determine the gauge-invariant fields.\cite{HCC}
In this work, we represented the resolvent field and the gauge field in the temporal gauge as
functions of spatial variables that are second-rank tensors 
in the combined spatial and SU(2) indices. Except in so far as 
the forms of $\overline{{\cal{A}}_{i}^{\gamma}}({\bf{r}})$ and 
$A_i^{\gamma}({\bf r})$ reflect this second-rank 
tensor structure, they are isotropic functions of position. In this way, we can represent the 
longitudinal part of the gauge field in the temporal gauge as  
 \begin{equation}
A_i^{\gamma}\,^L({\bf r})=\frac{1}{g}\left[{\delta}_{{i}\,{\gamma}}\,\frac{{\cal N}(r)}{r}+
\frac{r_i\,r_{\gamma}}{r}\,\left(\frac{{\cal N}(r)}{r}\right)^{\prime}\right]
\label{eq:longa}
\end{equation}
and the transverse part as
\begin{equation}
A_i^{\gamma}\,^T({\bf r})={\delta}_{{i}\,{\gamma}}\,{\cal T}_{A}(r)+
\frac{r_i\,r_{\gamma}}{r^2}\,{\cal T}_{B}(r)+{\epsilon}_{i{\gamma}n}\frac{r_n}{r}\,{\cal T}_{C}(r)
\label{eq:transa}
\end{equation}
where ${\cal N}(r)$, ${\cal T}_{A}(r)$, ${\cal T}_{B}(r)$ and ${\cal T}_{C}(r)$ are isotropic
functions of $r$, the prime denotes differentiation with respect to $r$, and 
the transversality of $A_i^{\gamma}\,^T({\bf r})$ requires that
\begin{equation}
\frac{d(r^2{\cal T}_B)}{dr}+r^2\frac{d\,{\cal T}_A}{dr}=0\,.
\label{eq:trans}
\end{equation}
An entirely analogous representation of the resolvent field enables us to relate it
to the gauge field through the nonlinear integral equation described in section \ref{sec:intro}. 
As a result of this analysis, we have
been able to show that it is possible to
represent the resolvent field as~\cite{HCC}
\begin{equation}
\overline{{\cal{A}}_{i}^{\gamma}}({\bf{r}})=\left({\delta}_{i\,{\gamma}}-\frac{r_i\,r_{\gamma}}{r^2}\right)
\left(\frac{\overline{\cal{N}}}{gr}+{\varphi}_A\right)+{\epsilon}_{i{\gamma}n}\frac{r_n}{r}\,{\varphi}_C
\label{eq:aform}
\end{equation} 
where 
\be
\varphi_A=\frac{1}{gr}\left[{\cal{N}}{\cos}(\overline{\cal{N}}+{\cal{N}})-{\sin}(\overline{\cal{N}}+
{\cal{N}})\right]
+{\cal T}_A\left[{\cos}(\overline{\cal{N}}+{\cal{N}})-1\right]-{\cal T}_C\,{\sin}(\overline{\cal{N}}+{\cal{N}})
\label{eq:phia}
\end{equation} 
and
\begin{equation}
\varphi_C=\frac{1}{gr}\left[{\cal{N}}{\sin}(\overline{\cal{N}}+{\cal{N}})+
{\cos}(\overline{\cal{N}}+{\cal{N}})-1\right]
+{\cal T}_C\left[{\cos}(\overline{\cal{N}}+{\cal{N}})-1\right]+{\cal T}_A\,{\sin}(\overline{\cal{N}}+{\cal{N}})
\label{eq:phic}
\end{equation}
\begin{equation}
\mbox{with}\;\;\overline{{\cal{Y}}^a}({\bf{r}})=
{\textstyle \frac{\partial_{j}}{\partial^{2}}
\overline{{\cal A}_{j}^{\alpha}}({\bf r})}=r_{a}\,\frac{\overline{\cal{N}}}{gr}\;\;\mbox{and}\;\;
{\cal{X}}^a({\bf{r}})={\textstyle\frac{\partial_i}{\partial^2}}A_i^\alpha({\bf{r}})=r_{a}\,\frac{\cal{N}}{gr}\,.
\label{eq:Yeq}
\end{equation}
Similarly, the gauge-invariant gauge field can be expressed as 
a functional of $\overline{\cal{N}}$ and of ${\cal{N}}$, ${\cal T}_A$, ${\cal T}_B$ and ${\cal T}_C$
as shown by  
\begin{eqnarray}
&&A_{{\sf GI}\,i}^{\gamma}({\bf{r}})=\frac{1}{gr}\left\{\epsilon_{i\,\gamma\,n}\frac{r_n}{r}
\left[{\cos}(\overline{\cal{N}}+{\cal{N}})-1+{\cal{N}}\sin(\overline{\cal{N}}+{\cal{N}})\right]
+\left({\delta}_{i\,\gamma}-\frac{r_ir_{\gamma}}{r^2}\right)\times\right.\nonumber\\
&&\left.\times\left[{\cal{N}}{\cos}(\overline{\cal{N}}+{\cal{N}})-{\sin}(\overline{\cal{N}}+{\cal{N}})\right]
-\frac{r_ir_{\gamma}}{r}\frac{d\overline{\cal{N}}}{dr}\right\}\nonumber \\
&&+{\cal T}_A\left\{
\left({\delta}_{i\,\gamma}-\frac{r_ir_{\gamma}}{r^2}\right)
{\cos}(\overline{\cal{N}}+{\cal{N}})
+\epsilon_{i\,\gamma\,n}\frac{r_n}{r}{\sin}(\overline{\cal{N}}+{\cal{N}})\right\}+\frac{r_ir_{\gamma}}{r^2}
\left({\cal T}_A+{\cal T}_B\right)\nonumber \\
&&+{\cal T}_C\left\{\epsilon_{i\,\gamma\,n}\frac{r_n}{r}{\cos}(\overline{\cal{N}}+{\cal{N}})
-\left({\delta}_{i\,\gamma}-\frac{r_ir_{\gamma}}{r^2}\right)
{\sin}(\overline{\cal{N}}+{\cal{N}})\right\}\,.
\label{eq:AGIsub}
\end{eqnarray}
With these representations, 
the nonlinear integral equation that relates the resolvent field to the gauge field 
is transformed to the nonlinear differential equation  
\begin{eqnarray}
\frac{d^2\,\overline{\cal{N}}}{du^2}&+&\frac{d\,\overline{\cal{N}}}{du}+
2\left[{\cal{N}}{\cos}(\overline{\cal{N}}+{\cal{N}})-
{\sin}(\overline{\cal{N}}+{\cal{N}})\right]\nonumber \\
&+&2gr_0\exp(u)\left\{{\cal T}_A\left[{\cos}(\overline{\cal{N}}+{\cal{N}})-1\right]-{\cal T}_C\,
{\sin}(\overline{\cal{N}}+{\cal{N}})\right\}=0
\label{eq:nueq}
\end{eqnarray}
where $u=\ln(r/r_0)$ and $r_0$ is an arbitrary constant. Eq. (\ref{eq:nueq}) 
relates\ $\overline{\cal{N}}$ to the source terms ${\cal{N}}$, ${\cal T}_A$ 
and ${\cal T}_C$. Together with these source terms, $\overline{\cal{N}}$ 
completely determines the resolvent field and the gauge-invariant gauge field.
In the limit in which the gauge field $A_i^a=0$, ({\em i. e.} ${\cal{N}}={\cal T}_A={\cal T}_B={\cal T}_C=0$),
Eq. (\ref{eq:nueq}) reduces to the autonomous ``Gribov equation''~\cite{gribov,gribovb}
\begin{equation}
\frac{d^2\,\overline{\cal{N}}}{du^2}+\frac{d\,\overline{\cal{N}}}{du}-
2{\sin}(\overline{\cal{N}})=0
\label{eq:nueqzero}
\end{equation}
 which is also the equation for a damped pendulum with $u$ representing the time, with the proviso
that $\overline{\cal{N}}$ must remain bounded not only in the interval $0{\leq}u<\infty$, but
in the larger interval $-{\infty}<u<\infty$ to include 
the entire space $0{\leq}r<\infty$.\sss

We have obtained numerical solutions of Eq. (\ref{eq:nueq}) with a variety of choices for the 
inhomogeneous terms ${\cal{N}}$, ${\cal T}_A$, ${\cal T}_B$ and ${\cal T}_C$. We have also defined 
the part of the gauge field 
\begin{equation}
{\sf A}\,_i({\bf{r}})=-ig\frac{{\tau}^{\gamma}}{2}\left[A_{{\sf GI}\,i}^{\gamma}({\bf{r}})\right]_{V}
=V_C({\bf{r}}){\partial}_iV^{-1}_C({\bf{r}})
\label{eq:Ufieldformal}
\end{equation} 
that has the structure (but not the physical significance) of a `` pure gauge'' field,  and have evaluated what we 
have called the ``winding number''~\cite{HCC}
\begin{equation} 
Q=-(24{\pi}^2)^{-1}{\epsilon}_{ijk}{\int}d{\bf{r}}\mbox{Tr}[{\sf A}\,_i({\bf{r}}){\sf A}\,_j({\bf{r}})
{\sf A}\,_k({\bf{r}})]\,.
\label{eq:Qformal}
\end{equation}
In evaluating $Q$, we have substituted Eq. (\ref{eq:Yeq}) into Eq. (\ref{eq:el1}), to obtain
\begin{equation}
V_C({\bf{r}})={\exp}\left(-i{\hat r}_n{\tau}_n\frac{\left(\overline{\cal{N}}+
{\cal{N}}\right)}{2}\right)\,.
\label{eq:VCY}
\end{equation}
In these numerical calculations, we assume that ${\cal T}_A$, ${\cal T}_B$ and ${\cal T}_C$ --- the 
transverse parts of the gauge field --- all vanish at the origin and as $r{\rightarrow}\infty$;
and that ${\cal{N}}$ vanishes at the origin and that, as $r{\rightarrow}\infty$, ${\cal{N}}{\rightarrow}2{\ell}\pi$ where 
$\ell$ is an integer, so that the pure-gauge field associated with the source field $A_i^a({\bf{r}})$,
$${\exp}\left(-i{\hat r}_n{\tau}_n\frac{{\cal{N}}}{2}\right)\partial_i\,{\exp}\left(i{\hat r}_n{\tau}_n
\frac{{\cal{N}}}{2}\right),$$
has an integer winding number.~\cite{Jack} On the basis 
of these numerical calculations, we have made the following observations:~\cite{HCC}\sss

\nin
${\bullet}$ Solutions of Eq. (\ref{eq:nueq}) exist that are bounded in the entire interval $0{\leq}r<\infty$
($-{\infty}<u<\infty$). When we normalize these solutions so that $\overline{\cal{N}}=0$ at $r=0$,
we find that, for cases in which $\ell=0$, 
$\overline{\cal{N}}{\rightarrow}m\pi$, where $m$ is an integer. Even values of $m$
correspond to positions of unstable equilibrium for the equivalent pendulum, and odd value of $m$ 
correspond to 
stable equilibrium positions. Numerical calculations that we have carried out never terminate with 
unstable equilibrium positions, because of rounding errors in the numerics. Both even and odd
values of $m$ are possible in the limit $r{\rightarrow}\infty$, even though numerical integrations only 
terminate with odd values of $m$ in  that limit. \sss

\nin
${\bullet}$ In the case of solutions of Eq. (\ref{eq:nueq}) for which 
$\lim_{r{\rightarrow}\infty}{\cal{N}}={\ell}\pi$,
with $\ell$ an integer other than $0$, $\lim_{r{\rightarrow}\infty}{\overline{\cal{N}}}={\tan}^{-1}(2{\pi}\ell)$.
\sss

\nin
${\bullet}$ The ``winding number'' $Q$, defined in Eq. (\ref{eq:Qformal}), has values $Q=m/2$ when $\ell=0$,
({\em i. e.} when $\lim_{r{\rightarrow}\infty}{\cal{N}}=0$), and is a half-integer when 
$m$ is odd, corresponding to a position of stable equilibrium in the $r{\rightarrow}\infty$ limit. When 
the integer $\ell{\ge}0$, 
\begin{equation}
Q=\frac{1}{2\pi}\left\{{\arctan}(2{\pi}{\ell})+m\pi+2{\pi}{\ell}
\left(1-\frac{1}{\sqrt{1+4{\pi}^2{\ell}^2}}\right)\right\}
\label{eq:windnew}
\end{equation}
where $m$ is an arbitrary integer and $\arctan$ is defined to represent the first 
sheet of the multivalued $\tan^{-1}$ function.\cite{jackpi} \sss

\nin
${\bullet}$ The nonlinear Eq. (\ref{eq:nueq}) has a number of different 
solutions, bounded in $0{\leq}r<\infty$, for the same set of source terms
${\cal{N}}$, ${\cal T}_A$, ${\cal T}_B$ and ${\cal T}_C$. Moreover, a number of numerical experiments 
have shown that the solutions of Eq. (\ref{eq:nueq}) are remarkably insensitive to these source 
terms. In particular, regardless of the value of $\lim_{r{\rightarrow}\infty}{{\cal{N}}}$, in the 
limit $r{\rightarrow}\infty$,  $A_{{\sf GI}\,i}^{\gamma}({\bf{r}})$ has the form of the hedgehog
\be
A_{{\sf GI}\,i}^{\gamma}({\bf{r}})=-\left(1{\pm}\sqrt{1+4\pi^2{\ell}^2}\right)\frac{1}{gr}\epsilon_{i\,\gamma\,n}\frac{r_n}{r}
\label{eq:mono}
\ee
where $\pm$ corresponds to negative (positive) values of 
$\sin\{\lim_{u{\rightarrow}{\infty}}\overline{\cal{N}}(u)\}$ and
$\cos\{\lim_{u{\rightarrow}{\infty}}\overline{\cal{N}}(u)\}$.

\section{Discussion}
The topological features of the resolvent field --- and of the gauge field --- that we have discussed in this 
work are not due to boundary values that we have imposed arbitrarily. They are 
direct consequences of the integral equation that establishes the resolvent field and that accounts 
for the implementation of Gauss's law and the gauge invariance of the fields described in 
Eqs. (\ref{eq:AdressedAxz})-(\ref{eq:psiqcdg1}). The topological features of the resolvent field and 
of the gauge-invariant gauge field that we have discussed in Section~\ref{sec:top}
therefore must be ascribed to the constraint imposed by the implementation of the non-Abelian Gauss's law
and by the {\em ansatz} chosen for the representation of the SU(2) gauge field and the resolvent field. \sss

Some of the observations made in Section \ref{sec:top} call for further investigations to clarify the 
relation between QCD dynamics and topology and between formulations of QCD in different gauges. For example, 
the observation that, for the same set of source terms, Eq. (\ref{eq:nueq}) has multiple
solutions, raises questions about the relation of these multiple
solutions to Gribov copies of gauge fields in 
the Coulomb gauge. Also, further work is required to fully understand how the ``hedgehog''
form of the gauge-invariant gauge field in the large-$r$ limit enables it to function as the
``glue'' in the nonlocal quark-quark and quark-gluon interactions in the representation of QCD
in which  the Hamiltonian  is represented entirely in terms of gauge-invariant fields.

\section*{Acknowledgments}
This research was supported by the Department of Energy
under Grant No. DE-FG02-92ER40716.00.

\end{document}